\documentclass[12pt]{iopart}
\usepackage{geometry,graphicx, iopams,float,subfigure}

\expandafter\let\csname equation*\endcsname\relax

\expandafter\let\csname endequation*\endcsname\relax
\usepackage{amsmath}

\begin{document}

\title{Chaos and ergodicity  in an entangled two-qubit Bohmian system }

\author{A.C. Tzemos and G. Contopoulos}

\address{Research Center for Astronomy and 
Applied Mathematics of the Academy of 
Athens - Soranou Efessiou 4, GR-11527 Athens, Greece}
\ead{thanasistzemos@gmail.com,
gcontop@academyofathens.gr}
\vspace{10pt}

\begin{abstract}
We  study in detail the onset of chaos and the probability measures formed by individual Bohmian trajectories in  entangled states of two-qubit systems for various degrees of entanglement.  The qubit systems consist of  coherent states of 1-d harmonic oscillators with irrational frequencies. In  weakly entangled states chaos is manifested  through the sudden jumps of the Bohmian trajectories between successive Lissajous-like figures. These jumps are succesfully interpreted by the `nodal point-X-point complex' mechanism. In strongly entangled states, the  chaotic form of the Bohmian trajectories is manifested after a short time. We then study the mixing properties of ensembles of Bohmian trajectories with initial conditions satisfying Born's rule. The trajectory points are initially  distributed in two sets $S_1$ and $S_2$ with disjoint supports but they exhibit, over the course of time, abrupt mixing whenever they encounter the nodal points of the wavefunction. Then a substantial fraction of trajectory points is exchanged between $S_1$ and $S_2$, without violating Born's rule. Finally, we provide strong numerical indications that, in this system, the main effect of the  entanglement is the establishment of ergodicity  in the individual Bohmian trajectories as $t\to\infty$: different initial conditions result to the same limiting distribution of trajectory points. 
\end{abstract}

\section{\label{sec:level1}Introduction}
Bohmian Quantum Mechanics (BQM) is an  interpretation of Quantum Mechanics
according to which the  quantum particles follow deterministic trajectories dictated by the Bohmian equations:
\begin{eqnarray}\label{bohmeq}
m_i\frac{dx_i}{dt}=\frac{\hbar}{G}\Big(\frac{\partial\Psi_I}
{\partial x_i}\Psi_R-\frac{\partial \Psi_R}
{\partial x_i}\Psi_I\Big),\quad i=1,2,\dots,
\end{eqnarray}
where $\Psi=\Psi_R+i\Psi_I$ is the wavefunction, guided by the usual time-dependent Schr\"{o}dinger equation $\hat{H}\Psi=i\hbar\partial \Psi/\partial t$, and $G=\Psi_R^2+\Psi_I^2$ \cite{Bohm, BohmII, holland1995quantum, deurr2009bohmian,benseny}.

Quantum Entanglement (QE) is a special physical property which characterizes the quantum systems and has a central role in the theory of Quantum Information and Computation, both from a theoretical and a technological standpoint. The construction of quantum computing machines is based solely on the understanding and manipulation of QE \cite{nielsen2004quantum, horodecki2009quantum}. In addition, QE is of fundamental importance in Statistical Mechanics of quantum systems, in general, since it governs the quantum probabilities of subsystems entangled with their environments.

The highly nonlinear character of the Bohmian equations  motivated several works studying the dynamics of Bohmian trajectories (see, e.g.   \cite{frisk1997properties, bialynicki2000motion, falsaperla2003motion, wisniacki2005motion, yang2008strong,  chou2008quantum, chou2009quantum, chou2009hydrodynamic}). In our previous works we studied the existence of order and chaos in 
Bohmian trajectories both in 2 and 3 dimensions 
and proposed a general theoretical framework describing 
the production of chaos in BQM \cite{ efthymiopoulos2006chaos, efthymiopoulos2007nodal, contopoulos2008order, efth2009, tzemos2018origin}. We also studied extreme 
cases where the Bohmian trajectories (both ordered and 
chaotic) evolve on certain integral surfaces, a phenomenon 
we called partial integrability \cite{tzemos2018integrals}. Furthermore, within the framework of BQM, the existence of chaos 
was found to be crucial for the convergence of the trajectory probability 
density to Born's rule.
\cite{efthymiopoulos2006chaos,efthymiopoulos2017chaos}.

Having as a goal the exploration of chaos in the presence of QE  (see also \cite{cesa2016chaotic, zander2018revisiting,elsayed2018entangled}), 
in a previous paper \cite{tzemos2019bohmian} we gave 
the basic characteristics of the Bohmian trajectories 
in an entangled two-qubit system, where the basic qubit 
states in one dimension are realized as properly engineered 
coherent states of the unperturbed quantum harmonic oscillator. 
We chose to work with this system, since it is quite well 
understood from a quantum mechanical standpoint and is easily 
constructed and controlled in the laboratory. Our 
study showed that this model, although simple, is 
characterized by  very rich dynamics in the Bohmian 
framework. We found that the choice of the physical 
parameters and especially of the frequency ratio $\omega_x/\omega_y$, 
can lead to chaotic motion (irrational $\omega_x/\omega_y$), 
periodic motion (rational $\omega_x/\omega_y$)  and even 
to integrable motion ($\omega_x/\omega_y=1$). 
This rich variety of Bohmian trajectories stems 
from two basic characteristics of our model: i) The high complexity of the 
Bohmian equations which appears despite the apparent simplicity of the underlying wavefunction, and
ii) the existence of an infinite number of nodal points of the wavefunction, forming a time-varying lattice structure \cite{tzemos2019bohmian}.
As already shown in 
\cite{efthymiopoulos2007nodal,efth2009,tzemos2018origin}, chaos
is introduced by the approach of a trajectory to a `nodal 
point-X-point complex' (NPXPC). This is a geometrical 
structure of the Bohmian flow which appears generically in the close neighbourhood of a moving nodal point.

In the present paper we study in detail the question of how chaos affects the way  the Bohmian trajectories produce a covering of the configuration space,  while satisfying Born's rule $P=|\Psi|^2$  at any time $t$. 

In the first part of the paper (Sections 2-4) we study in detail  the production of chaos in the Bohmian trajectories of this system through the prism of the NPXC mechanism, by monitoring the  motion of the multiple NPXPCs for different amounts of entanglement and we comment on the  short term evolution of the Bohmian trajectories.
In the second part (Section 5) we study the long term covering of the configuration space by  the Bohmian trajectories  where our main result is that, in this system, the increase of entanglement generates  approximate ergodicity after a moderately long time. In particular we study numerically individual Bohmian trajectories and find that the long term distribution of the points of an individual trajectory $(x(t),y(t))$ is independent from their initial conditions $(x(0),y(0))$. The convergence to the limiting distribution (corresponding to $t\to\infty$) can be reached in a much shorter time if we take a collective distribution of many initial conditions satisfying initially the Born rule.  Consequently the measure $D(x,y)$ formed by taking the average in time covering of the space by a single trajectory $(x(t), y(t))$ is approximately equal to the mean of the quantity $|\Psi|^2$. This is a remarkable property of our system, since it shows that entanglement implies not only chaos, but also ergodicity, a unique property of certain dynamical systems, which, to our knowledge, has not been addressed in Bohmian Mechanics. Finally in Section 6 we draw our conclusions and present some interesting questions for future research based on our results.

\section{Our model}
The time-dependent state $|\alpha(t)\rangle$ of a 1-d coherent state of the harmonic oscillator satisfies the eigenvalue equation
\begin{equation}
\hat{\alpha}|\alpha(t)\rangle=A(t)|\alpha(t)\rangle,
\end{equation} where $\hat{\alpha}$ is the anihilation operator and $A(t)=|A(t)|\exp(i\phi(t))$ its complex eigenvalue. The wavefunction $Y$ corresponding to the state $|\alpha(t)\rangle$  in  the position representation is given by:
\begin{equation}
Y(x,t)=\Bigg(\frac{m\omega}{\pi\hbar}\Bigg)^{\frac{1}{4}}
\exp\Bigg[-\frac{m\omega}{2\hbar}\Bigg(x-\sqrt{\frac{2\hbar}{m\omega}}
\Re[A(t)]\Bigg)^2+i\Bigg(\sqrt{\frac{2m\omega}{\hbar}}
\Im[A(t)]x+\xi(t)\Bigg)\Bigg],
\end{equation}\label{cs}
with
\begin{eqnarray}
\Re[A(t)]=a_0\cos(\sigma-\omega t),
\Im[A(t)]=a_0\sin(\sigma-\omega t)\\ \xi(t)=\frac{1}{2}\Big[a_0^2
\sin(2(\omega t-\sigma))-\omega t\Big],
\end{eqnarray}
where
$\sigma=\phi(0)$ is the initial phase of the complex eigenvalue $A(t)$ and $a_0\equiv |A(0)|$. Following our previous paper \cite{tzemos2019bohmian} we study  a system of two non interacting 1-d oscillators along  the $x$ and $y$ direction respectively \begin{eqnarray}
H=\frac{p_x^2}{2m_x}+\frac{p_y^2}{2m_y}+\frac{1}{2}m_x\omega_x^2x^2+\frac{1}{2}m_y
\omega_y^2y^2.
\end{eqnarray} 
The state of our system is described by wavefunctions of the form:
\begin{equation}\label{kym}
\Psi(x,y,t)=c_1Y_R(x,t)Y_L(y,t)+c_2Y_L(x,t)Y_R(y,t),
\end{equation}
where 
\begin{eqnarray}
\nonumber Y_R(i,t)\equiv Y(i,t;\omega=\omega_i,m=m_i,\sigma=\sigma_i),\,i=x,y\\
Y_L(i,t)\equiv Y(i,t;\omega=\omega_i,m=m_i,\sigma=\sigma_i+\pi), i=x,y.
\end{eqnarray}
$Y_R$ and $Y_L$ are one-dimensional coherent states with center started to  the right or to the left from the center of the oscillation along   $x$ or $y$. By choosing $a_0=5/2$  the wavefunctions $Y_R$ and $Y_L$ acquire a negligible overlap in Hilbert space, and can be realised as  the basis states for a qubit. Consequently the wavefunction $\Psi(x,y,t)$ refers to an entangled state of two  qubits made of coherent states along the coordinates $x$ and $y$. The entanglement depends on the values of the constants $c_1, c_2$, which are chosen here to be real for the sake of simplicity with $c_1^2+c_2^2=1$. Consequently we can control the amount of entaglement by altering the value of one of these amplitudes, e.g. $c_2$, with $c_2\in[0,\sqrt{2}/2]$, where $c_2=0$ corresponds to a product state and $c_2=\sqrt{2}/{2}$ corresponds a maximally entangled state (Bell state). We note that the absence of interacting terms in our Hamiltonian guarantees the conservation of QE.

\section{Nodal trajectories}
The positions of the nodal points of the wavefunction \eqref{kym} are the solutions of the equation $\Re{\Psi}=\Im{\Psi}=0$ and evolve in time according to:
\begin{eqnarray}
\label{xnod}&x_{nod}={\frac {\sqrt {2}
\left( k\pi\,\cos \left( 
\omega_{y}\,t \right) +\sin \left( 
\omega_{y}\,t \right) \ln  \left( 
\left| {\frac {c_{1}}{c_{2}}} \right|  
\right)  \right) }{4\sqrt {\omega_{x}}a_{0}\,\sin \left(  
\omega_{xy}  t \right) } }\\&
\label{ynod}y_{nod}={\frac {\sqrt {
2} \left(k\pi\, \cos \left( \omega_{x}t 
\right) +\sin \left( \omega_{x}t \right) 
\ln  \left(  \left| 
{\frac {c_{1}}{c_{2}}} \right|  \right)  
\right) }{4\sqrt {\omega_{y}}a_{0}\,\sin 
\left( \omega_{xy}\,t \right) }}
\end{eqnarray}
with $k\in Z $, $k$ even for $c_1c_2<0$
or odd for $c_1c_2>0$ and $\omega_{xy}\equiv \omega_x-\omega_y$. We choose to work with positive $c_1,c_2$, $\omega_x=1$, and $\omega_y=\sqrt{3}$. Consequently the ratio $\omega_x/\omega_y$ is irrational and the $k's$ are odd.
From Eqs.~(\ref{xnod},\ref{ynod}) we see that the distance between two nodal points labeled by the integers $k_1$ and $k_2$ is $d\propto |k_1-k_2|$. In particular the distance between the nodal points $k_i$ and $k_{i+1}$ is equal to that between $k_i$ and $k_{i-1}$ at every given time $t$. This implies that the nodal points form a lattice structure on the plane $(x,y)$ at any time $t$.

\begin{figure}
\centering
\includegraphics[scale=0.4]{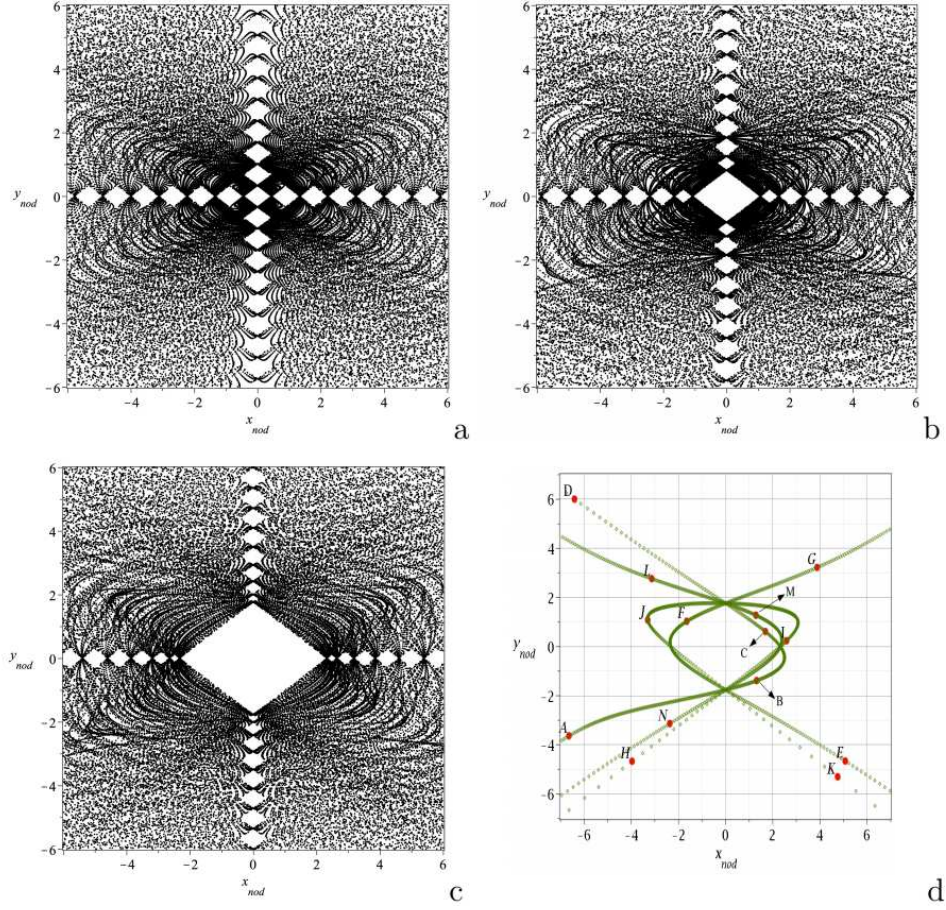}
\caption{Nodal trajectories corresponding 
to the odd values of $k\in[-23, 23]$ for $t\in[0,100]$ with 
fixed time step $t_s=0.01$ and different values of entanglement: 
(a) $c_2=\sqrt{2}/{2}$ (maximal entanglement)  
(b) $c_2=10^{-3}$ 
(c) $c_2=10^{-7}$. The 
density of the trajectories in the area around the origin  decreases
with the decrease of the entanglement (the central empty 
region increases as $c_2$ decreases). The areas of high 
concentration of points are those where the velocity of 
the nodes is relatively small. (d) 
The trajectory of the nodal point $k=1$ 
for $t\in[0,17.2]$ when $c_2=10^{-7}$. (a): The nodal point starts at 
$x = y = -\infty$ 
and passes through the points 
A-N at the times in parentheses: A(0.34), B(2.05), C(3.20), 
D(3.95), E(4.80), F(6.70), G(7.80), H(8.80), I(9.30), J(12.00), 
K(12.72), L(13.81), M(14.73), N(16.50).  Dots indicate large velocities.
 }
\label{npks}
\end{figure}

 The trajectories of different nodal points (for many odd values of $k$) for some values of entanglement are shown in Fig.\ref{npks}.
These trajectories fill the whole central part of the plane $(x,y)$
 except for the white regions. We observe that even with a 
 very small amount of entanglement $(c_2=10^{-7})$ the moving 
 nodal points cover a large part of space. This fact plays a crucial role in our analysis, as it implies that we have a significant probability of close encounters between nodal points and their surrounding nodal point-X-point complexes (NPXPC, see below) and the wandering Bohmian trajectories (see section 4), as $t\to\infty$. The decrease of entanglement enlarges the central empty region (the higher 
 the density in an area the lower the velocity of the nodes 
 in this area). Consequently, the increase of entanglement confines
 the ``slow part" of the nodal trajectories closer to the origin.
In the case of a product state $(c_2=0)$ we have 
$x_{nod}=y_{nod}=\infty$, i.e. there are no nodal points.

The study of the velocities of the nodal point is important because nodal points with large velocities  do not affect significantly the Bohmian trajectories, as pointed out in \cite{efth2009} (sections II C, D).

\section{Bohmian trajectories}

The Bohmian equations  of motion derived by substituting the wavefunction \eqref{kym} in  Eqs.~\eqref{bohmeq}  are:
\begin{eqnarray}
\nonumber &\frac{dx}{dt}=-\frac{\sqrt{2\omega_x}}{G}a_0\Big(A\cos(\omega_x t)+B\sin(\omega_x t)\Big)\\&
\frac{dy}{dt}=\frac{\sqrt{2\omega_y}}{G}a_0\Big(A\cos(\omega_y t)+B\sin(\omega_y t)\Big)\label{bohmI}
\end{eqnarray}
where 
\begin{eqnarray}
&A= 2c_{1}c_{2}{{\rm e}^{2
f_{x}+2f_{y}}}\sin \left( 2(g_{x}-g_{y})\right),\quad
B=c_1^2e^{4f_x}-c_2^2e^{4f_y}\\&
G=2c_{1}c_{2}{{\rm e}^{2f_{x}+2f_{y}}
}\cos \left( 2(g_{x}-g_{y}) \right) +{{\rm e}^{4f_{y}}}{c_{2}}^{2
}+{{\rm e}^{4f_{x}}}{c_{1}}^{2}
\end{eqnarray}
with
\begin{eqnarray}
\nonumber f_{x}=\sqrt {2\omega_{x}} a_{0}\cos \left( \omega_{x}
\,t \right)x,\quad f_{y}=\sqrt {2\omega_{y}}a_{
0}\cos \left( \omega_{y}\,t \right) y,\\
g_{x}=\sqrt {2\omega_{x}}a_{0}\,\sin \left( \omega_{x}\,t
\right) x,\quad g_{y}=\sqrt {2\omega_{y}}a_{0}\,\sin \left( \omega_{y}\,t
\right) y
\end{eqnarray}
and $c_1^2+c_2^2=1$.

In the case  $c_2=0$ (product state) the system acquires the solution:
$
\Big(x-x_0=\frac{\sqrt{2\omega_x}}{\omega_x}a_0(\cos(\omega_xt)-1),\,
y-y_0=-\frac{\sqrt{2\omega_y}}{\omega_y}a_0(\cos(\omega_yt)-1)\Big)
$
where $(x_0, y_0)$ are the initial conditions for $t=0$. Thus the trajectories are Lissajous figures, whose size is the same for all initial conditions $\Big(|\Delta x|=\frac{2\sqrt{2\omega_x}}{\omega_x}a_0\simeq7.1, |\Delta y|=\frac{2\sqrt{2\omega_y}}{\omega_y}a_0 \simeq5.4\Big)$.

\begin{figure}[hbt]
\centering
\includegraphics[scale=0.35]{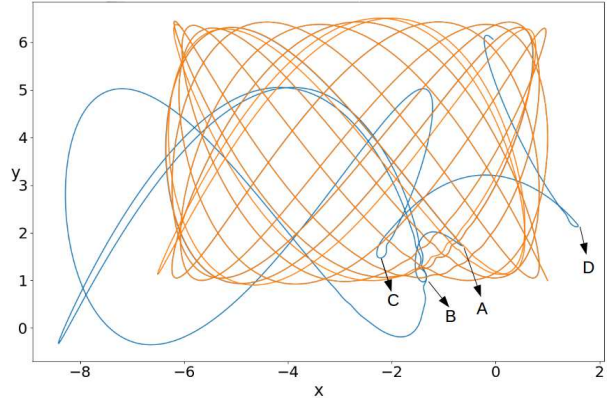}
\caption{Two trajectories up to $t = 60$, starting at ($x(0)=y(0)=1$) with $c_2=1.025\times10^{-7}$ (red) and $c_2=1.03$ (blue). The two trajectories almost coincide up to $t=43.3$ (point A). After that the red orbit continues without any major perturbation, while the blue orbit has abrupt perturbations (close approaches to a NPXPC) at points $A(t=43.3), B(t=44.1), C(t=57.3)$ and $D(t=58.6)$.}
\label{red_blue}
\end{figure}

As the nodal point passes an inifinite number of times through the region occupied by the trajectories inside the support of the wavefunction of the system (the region of the configuration space in which the value $|\Psi|^2$ is relatively high), there is a high probability of a close interaction between the nodal points and  the trajectories. We refer the reader to \cite{efthymiopoulos2007nodal,efth2009,tzemos2018origin} on
how this interaction takes place. Briefly, the nodal points form in their neighborhood a structure of the quantum flow called the `nodal point-X-point complex' (NPXPC). The trajectories exhibit hyperbolic scattering events at each close encounter with a NPXPC, and these encounters are responsible for the chaotic behavior of the trajectories.

We  calculated some trajectories starting at the same initial point $(x_0=y_0=1)$ for various values of $c_2$. For very small values of $c_2$ the trajectories are close to Lissajous curves for some time, and then the trajectories become chaotic due to a close approach to a nodal point and its associated X-point \cite{efthymiopoulos2007nodal, efth2009}. It is remarkable that a relatively fast transition from a near-Lissajous curve (for times up to $t=1000$) to chaotic trajectories happens for quite small values of $c_2$. In fact the trajectory for $c_2=1.025\times 10^{-7}$ is very similar to  a Lissajous figure with $c_2=0$,  while for $|c_2|=1.03\times 10^{-7}$ the trajectory is chaotic after a time $t=43.3$ (Fig. \ref{red_blue}), although the two trajectories almost coincide up to this time.

\begin{figure}[hbt]
\centering
\includegraphics[scale=0.24]{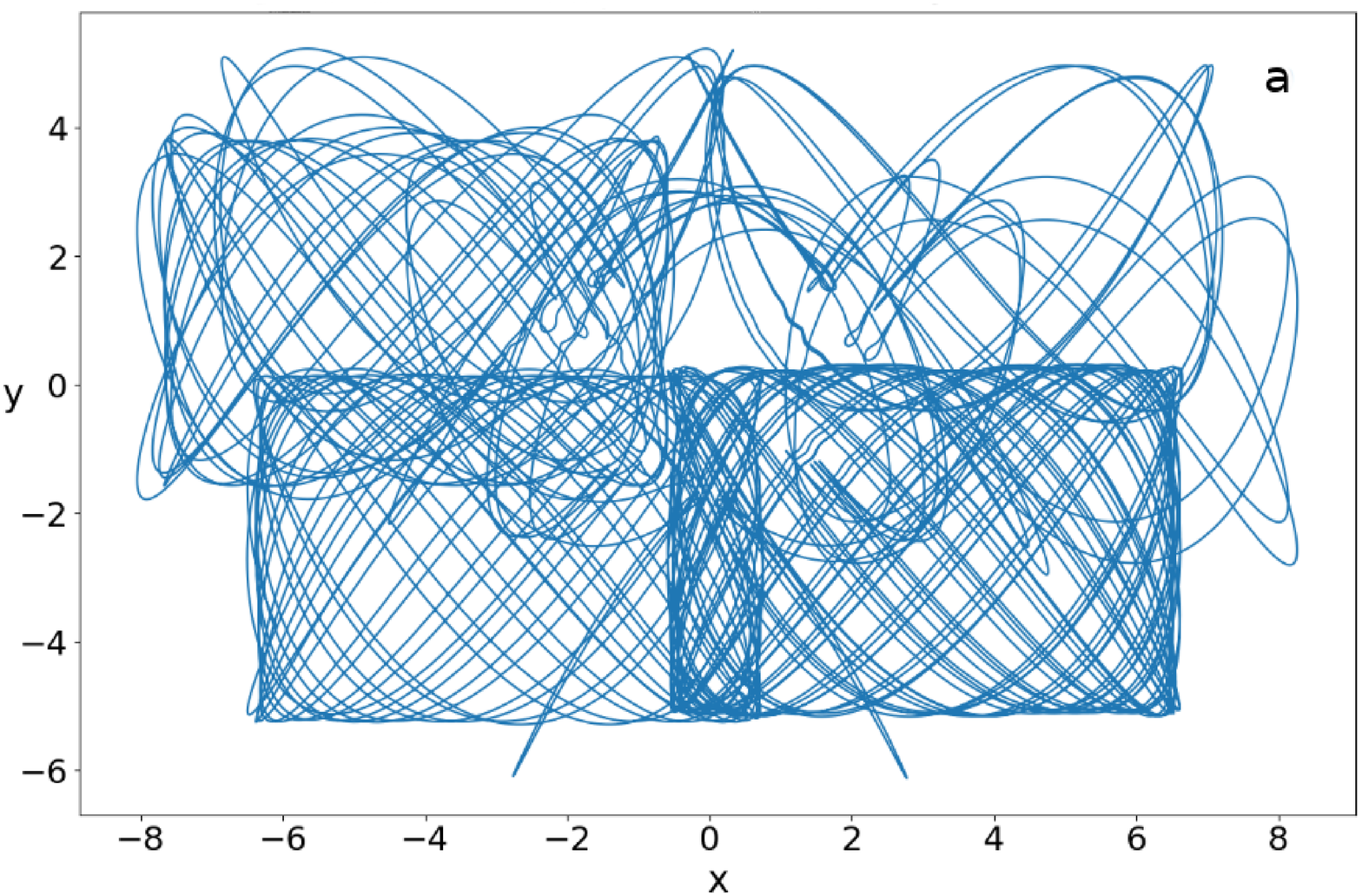}
\includegraphics[scale=0.28]{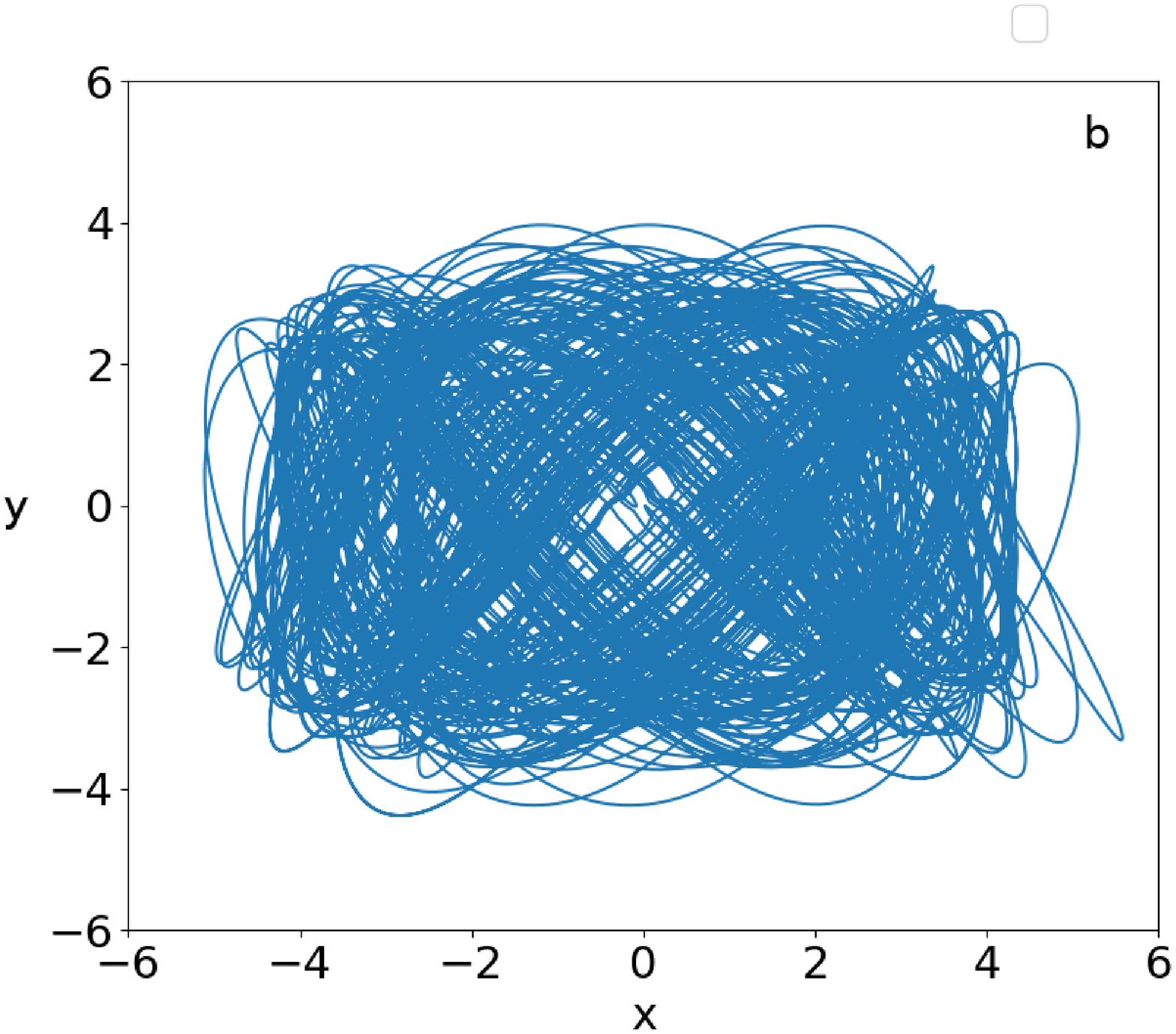}
\caption{ (a) A trajectory with $c_2=2\times 10^{-5}$ and $(x(0)=-2, y(0)=2)$ 
for time up to $t=500$. We observe the formation of Lissajous-like figures between the encounters of the trajectory with the NPXPCs. (b) A trajectory with the same initial conditions in the maximum entangled case $c_2 = \sqrt{2}/2 $ for a time up to $t=1000$. The motion is chaotic.}
\label{t175}
\end{figure}

When $c_2$ is small the orbit after a close approach to a NPXPC deviates from a Lissajous curve.  After  close approaches to NPXPCs it tends  to form several approximate Lissajous curves (Fig.~\ref{t175}a).  After a much longer time the orbit covers the central part of the configuration space corresponding to the support of the wavefunction, as we will see in the next session. On the other hand if $c_2$ is large there is no clear tendency to form approximate Lissajous curves and the trajectories are quite irregular from the beginning. An example is given in Fig.~\ref{t175}b for the case of maximum entanglement where $c_1=c_2=\sqrt{2}/2$.

\begin{figure}[H]
\centering
\includegraphics[width=7cm,height=6cm]{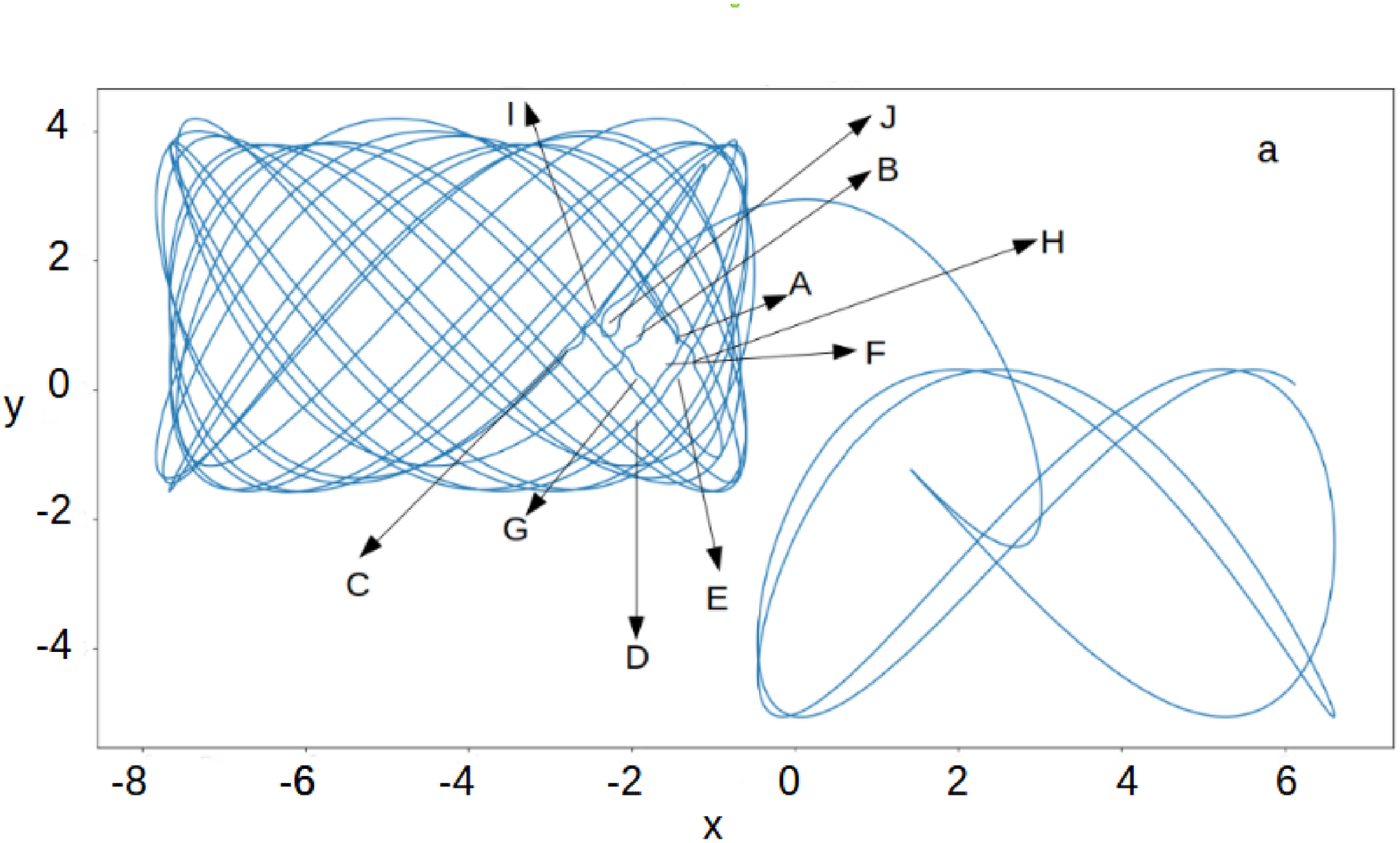}
\includegraphics[width=7cm,height=5.5cm]{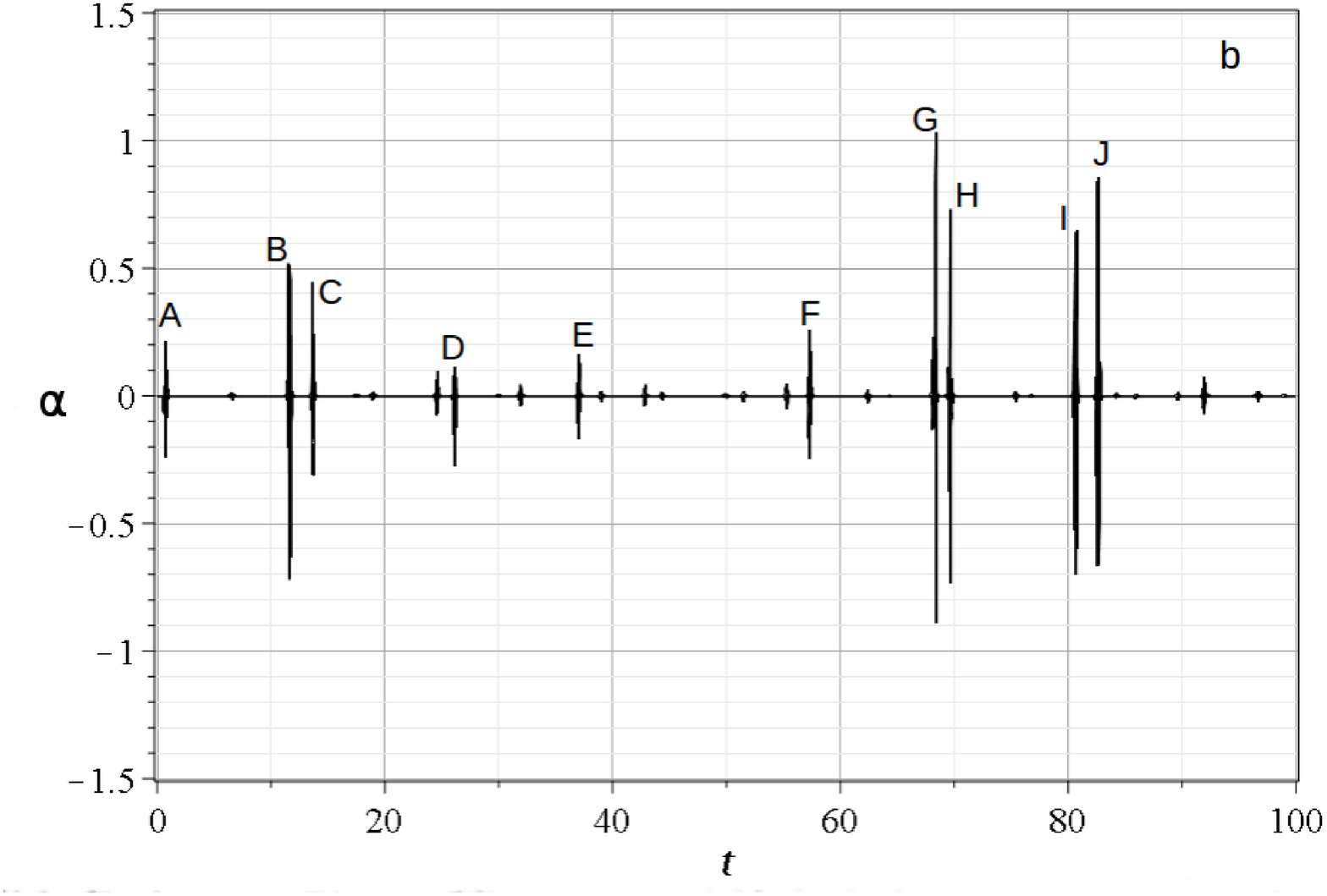}
\caption{
(a) An orbit with $c_2=2\times 10^{-5}$ and initial point ($x(0)=-2, y(0)=2$) for $t\in[0,100]$. The letters A-J indicate where the orbit approaches a nodal point-X-point complex at the times of fig.b. (b)
The stretching number $a$ has abrupt increases and decreases when the orbit approaches a nodal point at the times indicated in parentheses A(0.76), B(11.5), C(15.7), D(25.9), E(37.0), F(57.0), G(68.3), H(69.5), I(80.7), J(82.3).}
\label{letters}
\end{figure}

The effect of the approach of a trajectory NPXPC can be quantified in terms of the time behavior of the value of the stretching number
\begin{equation}
a_i=\ln\Big(\frac{\xi_{i+1}}{\xi_i}\Big),
\end{equation}
where $\xi_{i+1}/\xi_i$ are the infinitesimal deviations from a trajectory at two consequtive time steps $t_i, t_{i+1}$. In the case of the trajectory with $c_2=2\times 10^{-5}$ and initial conditions $x_0=-2, y_0=2$ (Fig.~\ref{letters}a) the main deviations of $a_i$ are marked (Fig.~\ref{letters}b) by successive letters A-J. The corresponding times are $t=0.76, 11.5\dots$ (as given in Fig.~\ref{letters}). At these times the trajectory is near the points A-J of Fig.~\ref{letters}a, where it exhibits some irregularity. At the points A and J the trajectory forms loops and reverses its direction.  
In fact at the point J the trajectory undergoes  large acceleration that brings it well outside the approximate Lissajous curve. One can verify that close to the other points (B-I) the trajectory shows also some small irregularities but does not form a loop. In fact, a detailed calculation shows that the loops observed at the points A and J correspond to close approaches in which the trajectory surrounds the nodal point. These are the so called `Type I scattering events' (Fig.~2a of \cite{efth2009}). On the other hand the other approaches (B-I) are of type II, as seen in the same figure  of \cite{efth2009}, and do not surround the nodal point.

\begin{figure}
\centering
\includegraphics[scale=0.4]{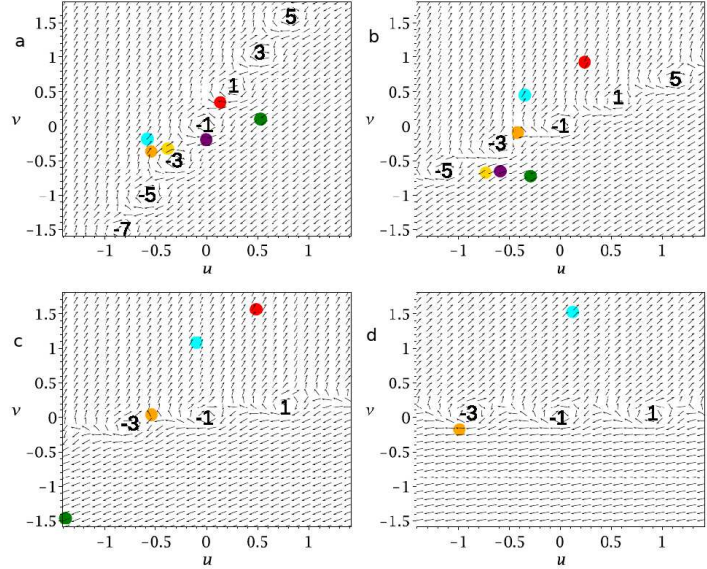}
\caption{The velocity field in a frame of coordinates centered at the nodal point $k=-1$ at particular times (a) $t=82.660$ (b) $t=82.845$ (c) $t=83.045 $ and $t=83.262$ for $c_2=2\times 10^{-5}$. The other nodal points and  X-points move around the nodal point $k=-1$. We give also the positions of 6 particles moving along the velocity field that are deflected as they approach some X-points. On the right of the center we have the nodal points $k=1, 3, 5, \dots$ and on the left the nodal points $k=-3,-5,\dots$}
\label{fields}
\end{figure}

In the present case we have an infinity of nodal points (and their associated NPXPCs) for various values of $k$. The trajectories may change their direction by approaching any one of these NPXPCs and they may go around any one of the nodal points. In Fig. \ref{fields} we have drawn the velocity field in a frame centered at the nodal point $k=-1$. The other nodal points are along a line which rotates slowly around the origin, along with a gradual in time change of the nearby velocity field.  The interaction between the trajectories and the nodal points is better illustrated by showing the Bohmian velocity field corresponding to the quantum flow as viewed in a frame of reference co-moving with one nodal point. Figure \ref{fields} shows the initial poisitions of six trajectory points, and their positions as time progresses are given in four snapshots.  We clearly see that the trajectory points follow the velocity field and are deflected as they approach some NPXPC.

In all the approaches the distance of the moving point from a nodal point, passes through a minimum. Furthermore, near the points A-J of Fig.~\ref{letters}a,b the velocity of the nodal point is close to a minimum. 

\section{Distributions}
If we consider solutions $\Psi$ of the Schr\"{o}dinger equation and distribute initially the  particles according  to  Born's rule
\begin{equation}\label{br}
P_0=|\Psi_0|^2
\end{equation}
at an initial time $t_0=0$, then the same relation applies at any later time \cite{valentini2005dynamical}. However it is of interest to see how this is realized in our model.

\begin{figure}[H]
\centering
\includegraphics[scale=0.35]{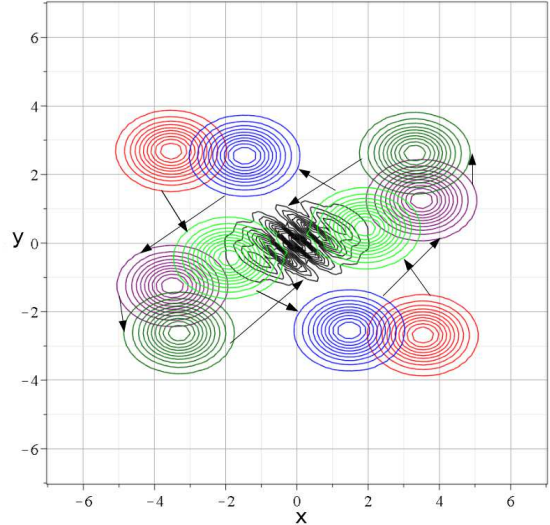}
\caption{The values of $|\Psi|^2$  in the case of maximum entanglement ($c_2 = c_1 = \sqrt{2}/2$) at  successive times: $ t=0$ (red), $t=1 $(green),$ t=2$ (blue), $t=3 $ (purple), $t=3.5$  (dark green) and $t\simeq 4.5$ (black).}
\label{support_motion}
\end{figure}

As a first example we find the solution of Schr\"{o}dinger equation for the maximally entangled case (Bell state \cite{nielsen2004quantum}) where $c_1=c_2=\sqrt{2}/2$. 
 In Fig.~\ref{support_motion} we see the values of $|\Psi|^2$  at successive times. Initially we have two symmetric blobs around the points $x_0=\pm3.54, y_0=\mp 2.68$ along the line with an inclination equal to $\theta\simeq-0.65$ radians (red contours). The two blobs have practically disjoint supports $S_1$ and $S_2$ in $x-y$ plane. Then at $t=1$ the blobs have moved closer to the center (green contours) and at $t=2$ they have passed to the other side (blue contours). Their motion is counterclockwise. At $t=3$ (purple contours) and at $t=4$ (dark green contours) they are close to an inclination $\theta\simeq 0.33$ but close to $t=4.5$ the two blobs collide and form a set of several maxima and minima. After this collision the blobs reappear and move around until a second collision takes place and so on.

\begin{figure}[H]
\centering
\hspace*{-0.8in}\includegraphics[scale=0.4]{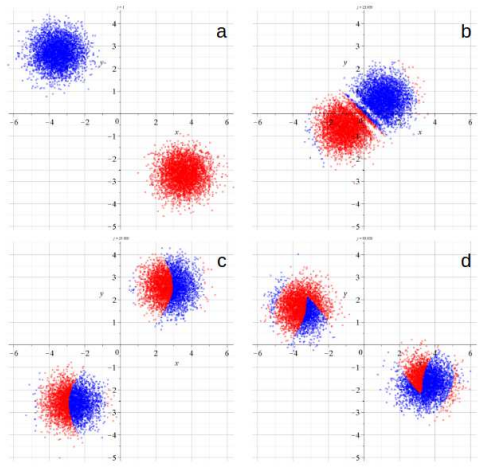}
\caption{The successive positions of two blobs containing 1000 initial conditions with distribution $P_0=|\Psi_0|^2$ in the case $c_1 = c_2$ at various times (a) $t=0$, (b) $t=4.5$ when the two blobs collide, (c) $t=5.8$ after the first collision when each blob is split, but their combination gives $P=|\Psi|^2$, (d) $t=16$ after the second collision (when again $P=|\Psi|^2$).}
\label{blobs}
\end{figure}

If now we populate the areas $S_1, S_2$ of the blobs with a large number of initial conditions distributed according to the law (\ref{br}), we can follow in detail how the ensemble of the resulting Bohmian trajectories evolves in time in order to always satisfy the law $P(t)=|\Psi(t)|^2$. Figure~\ref{blobs} provides the relevant information. The  trajectory points corresponding to initial conditions within  the left set $S_1$ are shown in blue and those of the initially right set $S_2$ are shown in red (Fig.~\ref{blobs}a). The domains covered by the blue and red points collide near $t=4.5$ as shown in Fig.~\ref{blobs}b. After the collision the domains covered by the red and blue points both split, so that the red points form two separate sets and the blue points form also two separate sets. However, the union of all the sets (red+blue) forms two blobs equal to the blobs of  $|\Psi|^2$ (Fig. \ref{blobs} c). Later on we have another collision and another splitting of the  red and  blue sets, but again the total distribution is very close to $|\Psi|^2$, as required by the theory (Fig.\ref{blobs} d). Even later we have further splittings but again the total density $P$ remains close to $|\Psi|^2$.

This result verifies in a spectacular way that while the relation $P(t)=|\Psi(t)|^2$ is satisfied at all later times if $P_0=|\Psi_0|^2$, the Bohmian trajectories undergo substantial mixing in the configuration space. The mixing is driven by the scattering of the trajectories in various directions, taking place at essentially every collision of the wavefunction blobs corresponding initially to $S_1$ and $S_2$.

\begin{figure}[H]
\centering
\includegraphics[scale=0.4]{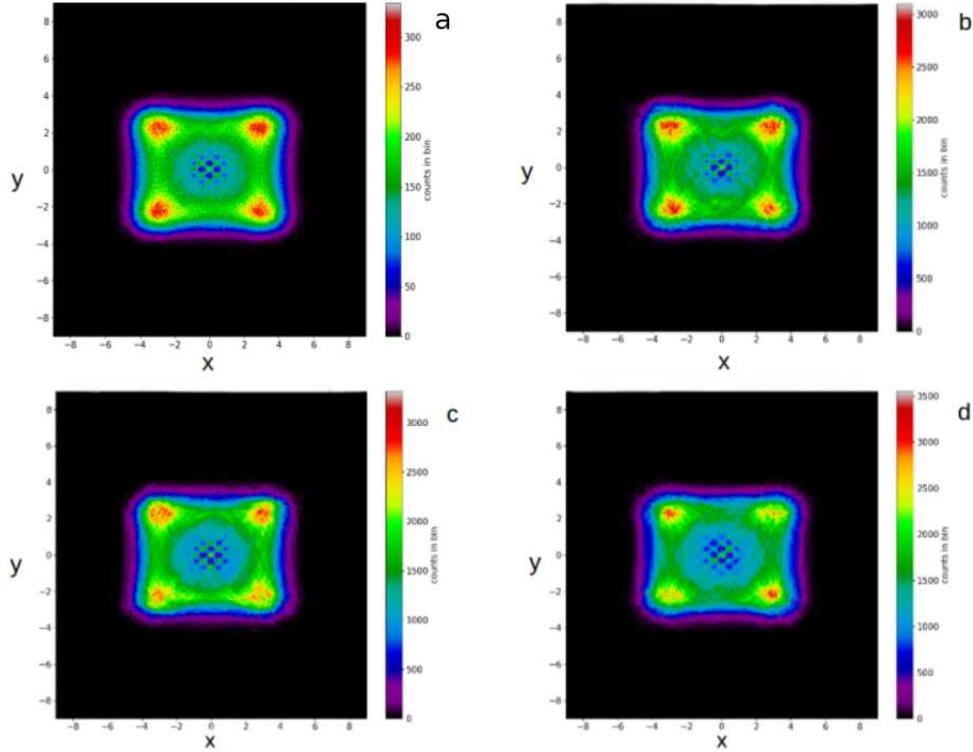}
\caption{(a) The overall distribution of 250 initial conditions approximating the formula $P_0=|\Psi_0|^2$  over a time  $ t=10^3$ (limiting distribution) . The distributions of the points of particular trajectories for $t$ up to $t=7\times 10^4$, given in Figs (b) for ($x_0=-2, y_0=2$), (c) for ($x_0=-4, y_0=4$) and (d) for ($x_0=-2, y_0=0$) approach very well the limiting distribution.}
\label{kare1}
\end{figure}

The overall coverage of the space by the red and blue 
points after a long time is shown in Fig.~\ref{kare1}a. 
This figure was calculated as follows. We covered  the central part of 
the configuration space $(x,y)$ by a large number of cells (a square grid of $360\times360$ cells) and we 
calculated the number of times the trajectories pass through 
these cells. The total number of initial conditions
was 250, distributed according to $P_0=|\Psi_0|^2$ 
(approximately) and the total time was $t=1000$. 
Then we gave to every cell  the number of passages of the trajectory  through this cell, by means of a color plot.

In a similar way, in Figs \ref{kare1}b,c,d we show individual trajectories with 
initial conditions $(x_0=-2, y_0=2)$, $(x_0=-4, y_0=4)$ and 
$(x_0=-2, y_0=0)$, calculated for a time $t=7\times10^4$ in 
order to have a significant coverage of space. All 
these figures have regions of large densities around 
the points $x=\pm 3.8, y=\pm 2.8$ and a very similar 
distribution of the regions with various colours, 
including the central regions where we see 4 islands, 
(surrounded by 8 islands) of smaller density 
than their surroundings.

The fact that these distributions are approximately 
the same with the total distribution of Fig.~\ref{kare1}a 
shows that individual trajectories tend to be ergodic. In fact, independently of the initial condition, all individual Bohmian trajectories have the same distribution after a long time. This property appears even while the system has no explicit outer boundary.  In particlular 
the trajectory of Fig.~\ref{kare1}c and a similar trajectory with 
initial conditions $(4,-4)$ start at points where 
$|\Psi|^2$ is very small at all times $t$.  We notice that the 
trajectories move to arbitrarily large distances only very rarely. This is indicated by the fact that 
the outermost parts of Figs~\ref{kare1} are dark 
purple, which means very small densities in the 
overall coverage. Nevertheless after a long time 
the trajectories that start far from the central region 
cover the space with approximately the same densities 
as the trajectories starting closer to the center.
Similar distributions are found for all the 
initial conditions of the Bell state (i.e. for $c_1=c_2=\sqrt{2}/2$). 

As $c_2$ decreases down to $c_2=0.01$ we found 
similar results, i.e. four blobs with maximum 
density but these maxima are closer to the 
center. E.g. while for $c_2=\sqrt{2}/2\simeq 0.707$ 
the maxima are at $x\simeq\pm 3.0, y\simeq\pm 2.4$, 
for $c_2=0.1$ the maxima are at 
$x\simeq\pm1.5, y\simeq\pm 1.3$ and for $c_2=0.01$ 
the maxima are at $x\simeq\pm 1.1, y\simeq\pm 0.7$. 

\begin{figure}[H]
\centering
\includegraphics[scale=0.15]{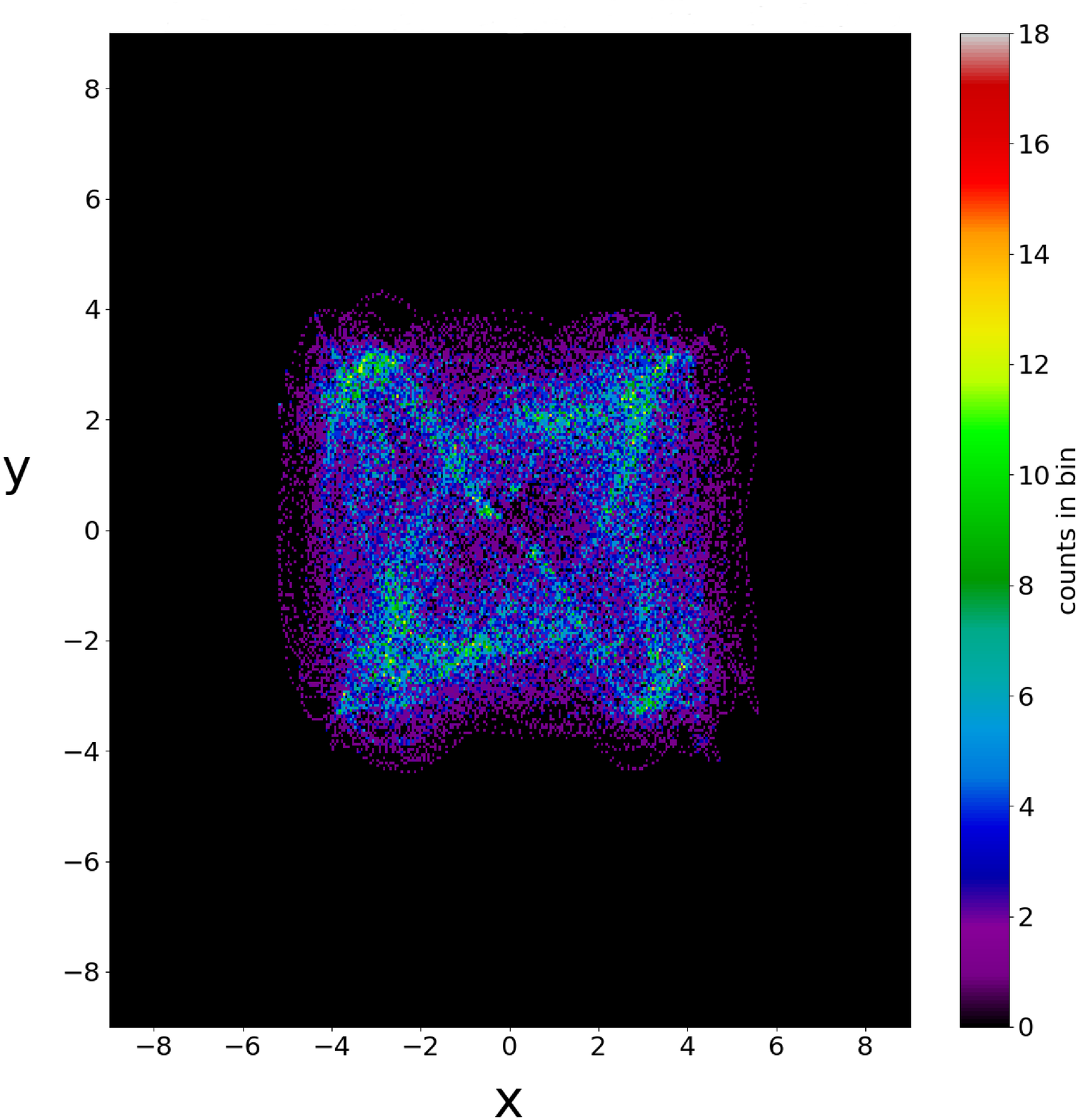}{a}
\includegraphics[scale=0.15]{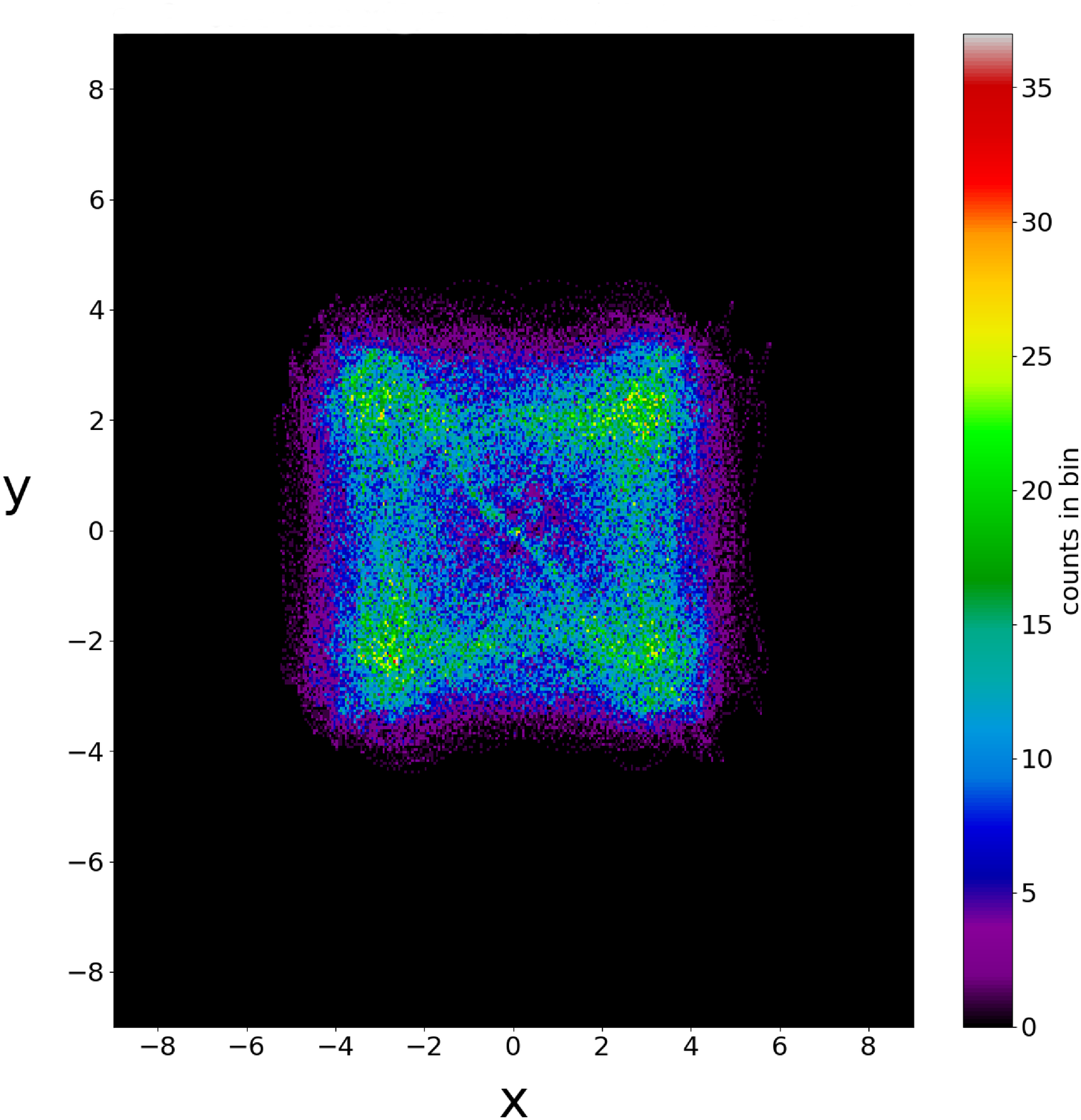}{b}
\caption{The distribution of the trajectory 
with initial conditions $x(0)=-2, y(0)=2$ in 
the case of maximal entanglement ($c_2\simeq 0.707$) 
for: (a) $t=4500$ and (b) $t=15000$: The trajectory 
acquires the characteristic form of a limiting 
distribution in a relatively short time.}
\label{bellp}
\end{figure}

\begin{figure}[H]
\centering
\includegraphics[width=0.4\textwidth]{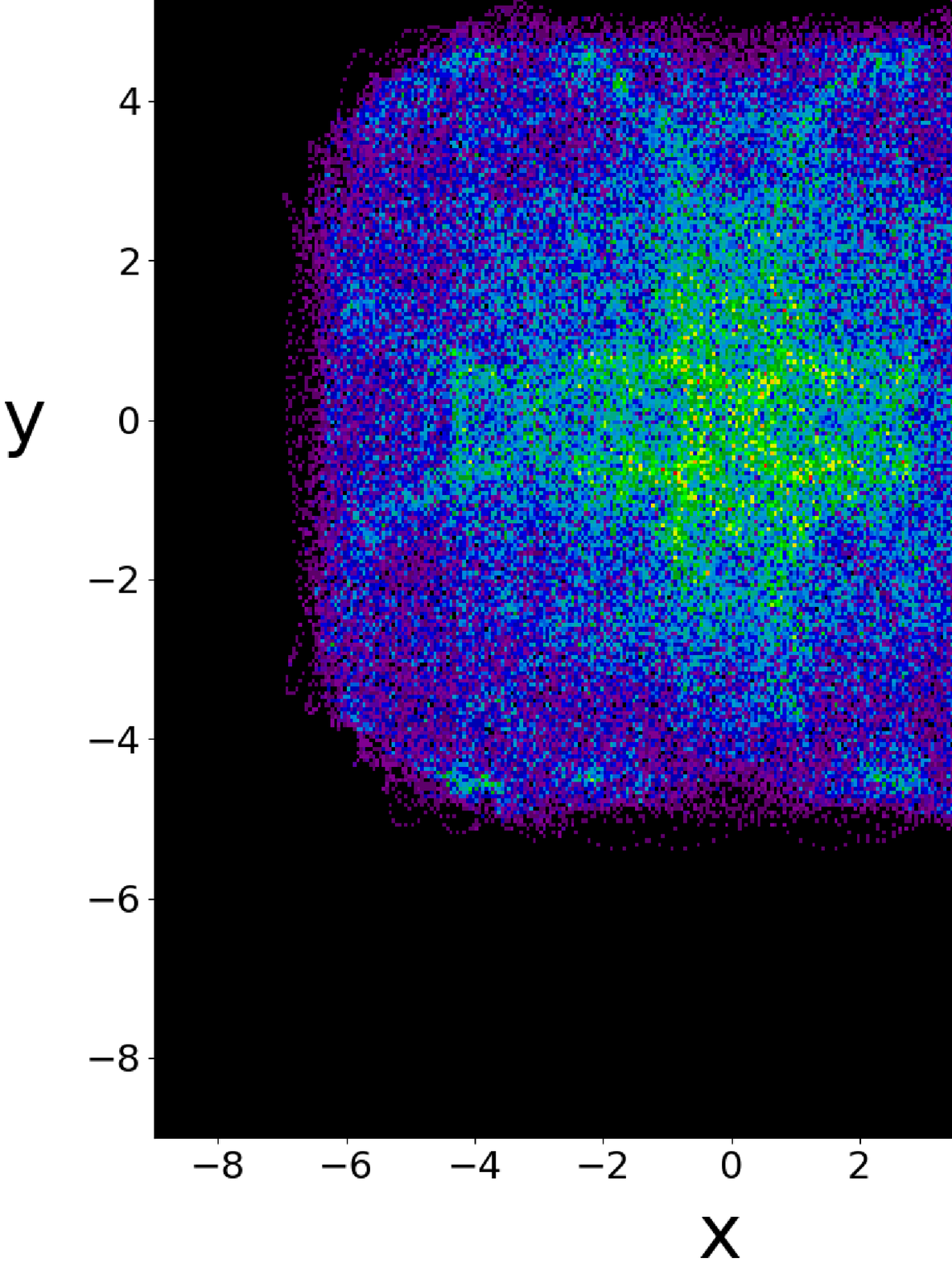}{a}
\includegraphics[width=0.4\textwidth]{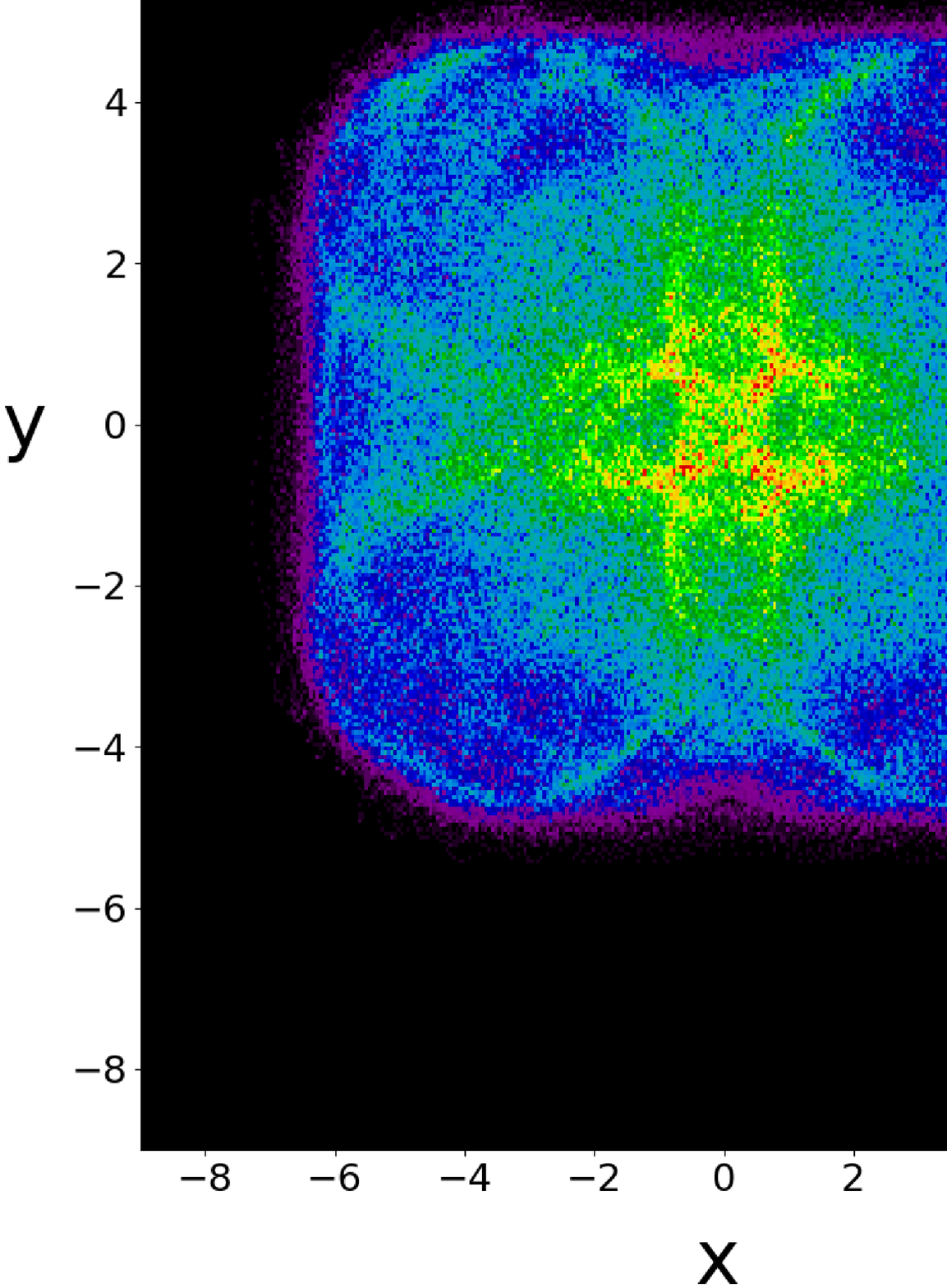}{b}
\caption{ The distributions of the points of a single Bohmian trajectory  
in the case of $c_2=0.01$, when  $x(0)=-2,\, y(0)=2$ for times (a) $t=15000$  and (b)  $t=75000$.}
\label{001}
\end{figure}

Most notably, the time needed by a trajectory to show 
its ergodic character (i.e. to approach the final 
average distribution) increases as $c_2$ decreases. 
Namely for $c_2=0.707$ we found that the main features 
of the limiting distribution of Fig.~\ref{kare1}a may 
be seen already for $t=15000$ (Fig.~\ref{bellp}). 
On the other hand for $c_2=0.01$ we need  a much longer time in order to approach the limiting
distribution. In Fig.~\ref{001}a we see that for $t=15000$ the distribution of the trajectory points does not represent clearly its final form. However, after a time $t=75000$ the distribution approaches satisfactorily its final form (Fig~\ref{001}b). We have found similar results for several trajectories and various values of $c_2$.

\section{Conclusions}

In this paper we continued our previous work \cite{tzemos2019bohmian}
on the effect of quantum entanglement in the evolution of Bohmian
trajectories, by studying in more detail and from various perspectives 
the trajectories of an  entangled two-qubit system. This simple bipartite 
system has the   advantage of being  simple from a computational point 
of view. However its unique feature, namely the infinite number of nodal 
points, provides us with rich information about the effect 
of entanglement on the Bohmian trajectories.

Our main conclusions are the following:

\begin{enumerate}
\item Chaos is introduced when the moving particles 
approach the nodal points whose velocities are small 
in aggreement with the theory of the NPXPCs.
\item In our model there are infinite nodal points that
fill most of the configuration space. Between the nodal points there 
are the X-points (unstable hyperbolic points of the flow). 
The distance between any two successive nodal points is 
the same for any given time $t$. The evolution of the nodal 
points has slow and fast parts: the fast parts correspond to the 
almost instantaneous departure   of  the NPXPCs from infinity along the diagonal 
lines and the slow parts correspond to the loops of the NPXPCs close 
to the origin, where the probability of strong scatttering events 
is high. 
\item For a very small amount of entanglement 
the Bohmian trajectories are of Lissajous type close
to the unperturbed Lissajous curves (of zero entanglement) 
 for very 
long times. For a somewhat larger  entanglement 
the Bohmian trajectories are again of Lissajous type for some time but they  are 
displaced with respect to the unperturbed Lissajous 
figures. After a close approach to a NPXPc these 
trajectories deviate considerably and form new Lissajous-type 
curves further away from the initial ones. Finally the 
trajectories fill chaotically the central region of the 
configuration space.
\item The initial form of the measure of the 
wavefunction $|\Psi_0|^2$
consists of two blobs. These blobs move around 
and from time to time 
they collide, but after every collision they 
form again two moving blobs. If
the initial conditions are distributed according to the probability law $P_0=|\Psi_0|^2$ then after every collision 
the trajectories of every blob split and follow 
very different directions. Nevertheless the total
distribution produced by the trajectory points always satisfies 
$P=|\Psi|^2$ (Born's rule).
\item For large entanglements the trajectories are 
quite chaotic. Their distribution tends to form 
very characteristic patterns. The individual trajectories, 
after a sufficiently long time tend to form the same 
patterns. Thus the trajectories tend to be ergodic although 
their initial conditions may be quite different. 
This is an important property that was numerically verified 
for large enough entanglements. We have numerical indications that the 
trajectories are ergodic in general, but the time needed 
for an orbit to show its ergodic character  
(approach to the average characteristic pattern) 
is  larger for smaller values of entanglement.  The time needed for the trajectories to converge to an ergodic behavior increases in general with the decrease of entanglement. Our present computational limitation allowed us to explore the domain $0.01\leq c_2\leq\sqrt{2}/2$. Further simulations  are required to explore the limit $c\to 0$.
\end{enumerate}

 Our main new result, the emergence of ergodicity in entangled systems raises 
new questions about the interplay between chaos and entanglement, 
which is an important problem in BQM. These first numerical 
resuts will be the starting point for future studies
on this subject addressing questions such as the 
ergodicity  of Bohmian particles a) in the presence of  weak entanglement, where 
we one expects to find
initial conditions which lead to ordered trajectories and b) in the 
special case when  
Born's rule is not initially satisfied. Finally an open problem is the relation between the degree of entanglement and the time needed to approach ergodicity.

\section{Appendix}

The existence of an infinite number of NPXPCs 
makes the integration of Bohmian trajectories a 
quite demanding process. 
During our calculations
we worked with different numerical methods of fixed 
and adaptive step size and we compared  their 
results. However there were many cases where the calculations
were impossible to finish in a reasonable amount of time, due 
to the extremely small stepsize needed to 
monitor the evolution of the trajectories (e.g. when the trajectory 
forms spirals around a moving nodal point)
in the case of fixed step methods, or due to 
the quick violation of 
error tolerance (smaller than $10^{-6}$) in 
the case of adaptive methods.

In order to observe these tiny details in the first moments of
the evolution and to have reliable results for large times
(in the case of distributions), we worked in Python 2.7 
with the  Lsoda adaptive numerical integration 
scheme \cite{linge2016programming}, 
with absolute error tolerance
$a_{tol}=10^{-10}$ and relative error tolerance 
$r_{tol}=10^{-7}$. Stricter error tolerance values did not alter our results.
\ack{The authors   thank Dr. C. 
Efthymiopoulos for his useful comments. This research is supported 
by the Research Commitee of the Academy of Athens. }

\section*{References}
\bibliographystyle{iopart-num}
\bibliography{bibliography}

\end{document}